\documentclass[aps,prb,a4paper,twocolumn,10pt,superscriptaddress,notitlepage]{revtex4-2}

\newcommand{\papertitle}{Launching Focused and Spatially Confined Phonon-Polaritons in Hexagonal Boron Nitride}

\usepackage[utf8]{inputenc}
\usepackage{graphicx}
\usepackage{wrapfig}
\usepackage{amsmath}
\usepackage{amssymb}
\usepackage{natbib}
\usepackage{float}
\usepackage{tikz}
\usepackage{siunitx}
\usepackage{dcolumn}
\usepackage{lmodern}
\usepackage[T1]{fontenc}
\usepackage[unicode=true,
	    colorlinks=true,
	    linkcolor=blue,
	    citecolor=blue,
	    urlcolor=blue]{hyperref} 
\usepackage{sansmath}
\usepackage{blindtext}
\usepackage[normalem]{ulem}

\usepackage{chemformula}
\usepackage{empheq}
\usepackage{comment}
\usepackage{bm}
\usepackage{gensymb}

\usepackage[misc]{ifsym}
\usepackage{fontawesome}

\usepackage{lineno}

\setcitestyle{super}

\usepackage{xr}
\makeatletter

\newcommand*{\addFileDependency}[1]{
\typeout{(#1)}
\@addtofilelist{#1}
\IfFileExists{#1}{}{\typeout{No file #1.}}
}\makeatother

\newcommand*{\myexternaldocument}[1]{%
\externaldocument{#1}%
\addFileDependency{#1.tex}%
\addFileDependency{#1.aux}%
}

\myexternaldocument{supplement}

\usepackage[top=2.5cm,bottom=2.5cm,left=1.5cm,right=1.5cm]{geometry}

\begin{document}

\title{\Large\textsf{\papertitle}}

\author{Bogdan Borodin}
\email{bborodin@nd.edu}
\affiliation{\footnotesize Department of Physics and Astronomy, University of Notre Dame, Notre Dame, IN 46556,~USA}
\affiliation{\footnotesize Stavropoulos Center for Complex Quantum Matter, University of Notre Dame, Notre Dame, IN 46556,~USA}

\author{Sergey Lepeshov}
\affiliation{Department of Electrical and Photonics Engineering, DTU Electro, Technical University of Denmark, Building 343, DK-2800 Kgs.~Lyngby, Denmark.}
\affiliation{NanoPhoton - Center for Nanophotonics, Technical University of Denmark, Ørsteds Plads 345A, DK-2800 Kgs.~Lyngby, Denmark.}

\author{Kenji Watanabe}
\affiliation{\footnotesize Research Center for Electronic and Optical Materials, National Institute for Materials Science, 1-1 Namiki, Tsukuba 305-0044,~Japan}

\author{Takashi Taniguchi}
\affiliation{\footnotesize Research Center for Materials Nanoarchitectonics, National Institute for Materials Science,  1-1 Namiki, Tsukuba 305-0044,~Japan}

\author{Petr Stepanov}
\email{pstepano@nd.edu}
\affiliation{\footnotesize Department of Physics and Astronomy, University of Notre Dame, Notre Dame, IN 46556,~USA}
\affiliation{\footnotesize Stavropoulos Center for Complex Quantum Matter, University of Notre Dame, Notre Dame, IN 46556,~USA}



\begin{abstract}
\vspace*{0.2cm}

Phonon-polaritons offer significant opportunities for low-loss, subdiffractional light guiding at the nanoscale. Despite extensive efforts to enhance control in polaritonic media, focused and spatially confined phonon-polariton waves have only been realized in in-plane-anisotropic crystals (e.g., MoO$_3$) and remain elusive in in-plane-isotropic materials (e.g., hexagonal boron nitride, hBN). In this study, we introduce a novel approach to launching phonon-polaritons by leveraging hBN subwavelength cavities at the Au/SiO$_2$ interface, enabling efficient coupling of cavities to the far-field component of mid-infrared light. Utilizing standard lithographic techniques, we fabricated subwavelength cavities of various shapes and sizes, demonstrating strong field enhancement, resonant mode localization, and generation of propagating phonon-polaritons with well-defined spatial structure. The cavity geometry governs wavefront curvature, spatial confinement, and polariton focusing, providing control over their propagation and achieving record-high in-plane confinement up to $\lambda/70$. Scattering-type scanning near-field optical microscopy reveals the real-space optical contrast of these cavity-launched modes, allowing for detailed characterization. We believe that our cavity-based approach to phonon-polariton focusing in isotropic media will pave the way for advanced nanophotonic applications.

\end{abstract}


\maketitle

\textbf{Introduction.} 
Strong light-matter interactions play a crucial role in well-established photonic devices, such as lasers, integrated photonic circuits, sensors, and hold promise for advancing future photonic quantum technologies, single-photon sources, entangled spin-photon interfaces, and quantum memories\cite{rivera2020light}. In particular, strong light-matter interactions can be achieved in propagating and localized phonon-polaritons (PhPs), which result from coupling between photons and phonons, quantized vibrations of a crystal lattice. PhP resonances occur in various crystalline materials within Reststrahlen band -- a region of the spectrum between transverse (TO) and longitudinal (LO) optical phonons, where the materials exhibit properties of a hyperbolic medium\cite{poddubny2013hyperbolic}. PhPs have garnered significant attention as they confine light substantially below the diffraction limit\cite{henry1965raman, caldwell2014sub, li2021direct}. This enables PhP waveguides and cavities with exceptional electromagnetic confinement in the mid-infrared range (mid-IR) and with remarkably low propagation loss compared to plasmon-polaritons~\cite{giles2018ultralow}. The mid-IR range of the spectrum is highly sought after due to its potential applications in medical diagnostics \cite{waynant2001mid, petrich2001mid}, bio- and gas sensing \cite{de2018applications, popa2019towards, lambert2015differential}, LiDaRs for spacecraft \cite{degtiarev2006compact,groussin2016thermap,olsen2020first}, optomechanics \cite{chen2021continuous}, and optical trapping \cite{zhang2016towards}. To achieve micrometer-scale integration of devices in the mid-IR range, the strong confinement provided by hyperbolic materials is likely the only viable solution.

\begin{figure*}[t]
    \centering
    \includegraphics[scale=0.95]{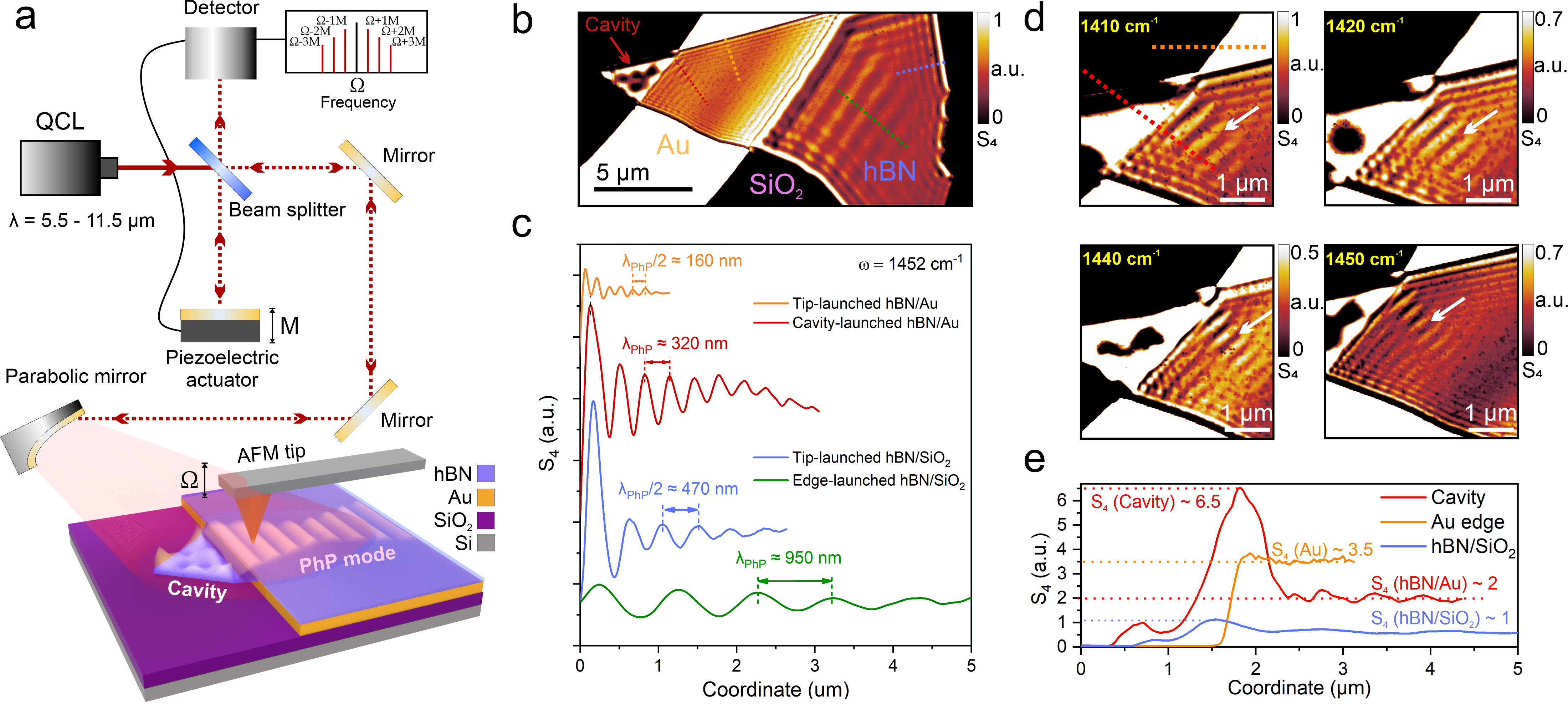}
    \caption{
    	\textbf{Diversity of PhP waves}. \textbf{a}) Schematics of the setup for the s-SNOM measurements of PhPs in hBN sub-wavelength resonators. Note that the far-field light covers the majority of the sample's area (spot radius $\geq$ 10 $\mathrm{\mu}$m) \textbf{b}) Optical amplitude (S$_4$) image of the 42-nm-thick hBN flake placed on the pre-patterned Si/SiO$_2$ substrate with Au stripes. Colored dotted lines indicate line-cuts of various propogating PhP waves. \textbf{c}) S$_4$ profiles taken across corresponding propagating PhP waves. \textbf{d}) S-SNOM image of the cavity and propagating PhPs at various excitation wavelengths. \textbf{e}) S$_4$ profiles taken across the Au stripe and the cavity at $\omega$ = 1410 cm$^{-1}$.}
     \label{fig1}
\end{figure*}

Hexagonal boron nitride (hBN) is one of the most promising and technologically accessible materials that provides a hyperbolic optical response in the mid-IR region of the spectrum. hBN is a uniaxial hyperbolic material that exhibits two types of hyperbolic optically active transitions — Type I - 760-820 cm$^{-1}$ ($\varepsilon_{xx} = \varepsilon_{yy} > 0 ; \varepsilon_{zz}<0$) and Type II - 1360-1600 cm$^{-1}$ ($\varepsilon_{xx} = \varepsilon_{yy} <0 ;\varepsilon_{zz} > 0$) \cite{caldwell2014sub,duan2017launching}, where $\varepsilon_{ij}$ is the permittivity tensor components. Recent studies have reported on the reliable growth of large-scale high-quality crystals with the possibility to control their isotopic purity \cite{wang2019epitaxial,liu2018single,kubota2007deep}. Currently, antennas are widely used to couple the far-field component of light with polaritonic systems, facilitating the selective launch of phonon-polaritons (PhPs)\cite{ni2021long,pons2019launching,li2015hyperbolic,dai2015subdiffractional}. However, this approach suffers from two critical limitations: the lack of control over the wavefront and low in-plane spatial confinement of the launched PhPs. While hBN nanocavities exhibit remarkable mid-IR light confinement down to $\lambda$/500, where $\lambda$ is the incident wavelength, an extraordinary Purcell factor (up to 10$^{12}$), and a high quality factor ($Q > 400$), the resonant modes in these cavities remain trapped within the resonators\cite{guo2023hyperbolic,tamagnone2018ultra,alfaro2017nanoimaging,herzig2024high}. Up to our best knowledge, there has been no demonstration of utilizing these localized modes to focus propagating PhPs, although PhP focusing along the optical axis has been achieved in biaxial hyperbolic materials such as MoO$_3$~\cite{zheng2022controlling}, where the focusing is fundamentally constrained by crystallographic directions, making it unsuitable for arbitrary-lens applications. This poses a significant limitation on the advancement of PhP-based technologies, and the effective control over the spatially-confined propagation and focusing in uniaxial hyperbolic materials remains elusive. 

Here, we demonstrate focused, spatially-confined, and long-lived PhPs that are launched by carefully engineered subwavelenght cavities, fabricated using standard lithographic protocols. We fabricate sub-wavelength hBN cavities at the edges between Au and Si$^{++}$/SiO$_2$ (285 nm) substrates (see Fig.~\ref{fig1}a). The Au edge acts as a scatterer, compensating for the momentum mismatch between incident light and PhPs. The Au/SiO$_2$ interface enables the coupling of PhPs with the out-of-plane far-field component of the incident light and serves as an in-plane partially transparent mirror for PhPs. On the SiO$_2$ side, it localizes a mode within the cavity, while on the Au side, it facilitates the launch of propagating and focused PhPs into the hBN bulk. Using scattering-type scanning near-field optical microscopy (s-SNOM), we visualize the real-space optical contrast and distinguish between cavity-launched ($\lambda_p$) and tip-launched ($\lambda_p$/2) modes. The cavity-launched PhPs result from coupling of the incident mid-IR light to the cavity modes and exhibit significant shape-dependent focusing. This approach for launching PhPs contrasts with s-SNOM tip-mediated excitation, which is employed in previous demonstrations of focused PhPs~\cite{chaudhary2019polariton}. We propose a new method for launching and focusing spatially confined in-plane PhPs, fully compatible with standard lithographic procedures. By fabricating cavities with various shapes, that determine the focal points and wavefronts of the PhPs, we demonstrate "concave lens"-like and "convex lens"-like focusing. Concave lens"-like cavities achieve confinement down to $\lambda/70$ at the focal point, setting a record-high confinement compared to previous studies on in-plane-focused PhPs\cite{qu2022tunable, zheng2022controlling,ma2024plane,liang2024manipulation}.

\vspace{0.2cm}
\textbf{Results.} In this work, we study two types of samples: (i) mechanically cleaved hBN crystals (of 30-45 nm thickness) transferred on the pre-patterned Au on Si$^{++}$/SiO$_2$ (285 nm) substrates as is (see Methods) and (ii) cavities fabricated at the interfaces between SiO$_2$ and Au using electron-beam lithography and plasma etching (see the schematics in Fig.\ref{fig1}). Fig.~\ref{fig1}a shows the measurement scheme for s-SNOM accompanied by the schematics of the fabricated sample. Here, we use a commercially available s-SNOM equipped with PtIr-coated (23 nm thick) AFM tips with the curvature radii less than 25 nm. A tunable ($\lambda$ = 5.5-11.5 $\mu$m) mid-infrared quantum cascade laser allows us to cover Type II hyperbolic transition - 1360-1600 cm$^{-1}$ ($\epsilon_{xx} = \epsilon_{yy} <0;\epsilon_{zz} > 0$) (see Methods section for details). 

\begin{figure*}[t]
    \centering
    \includegraphics[scale=0.9]{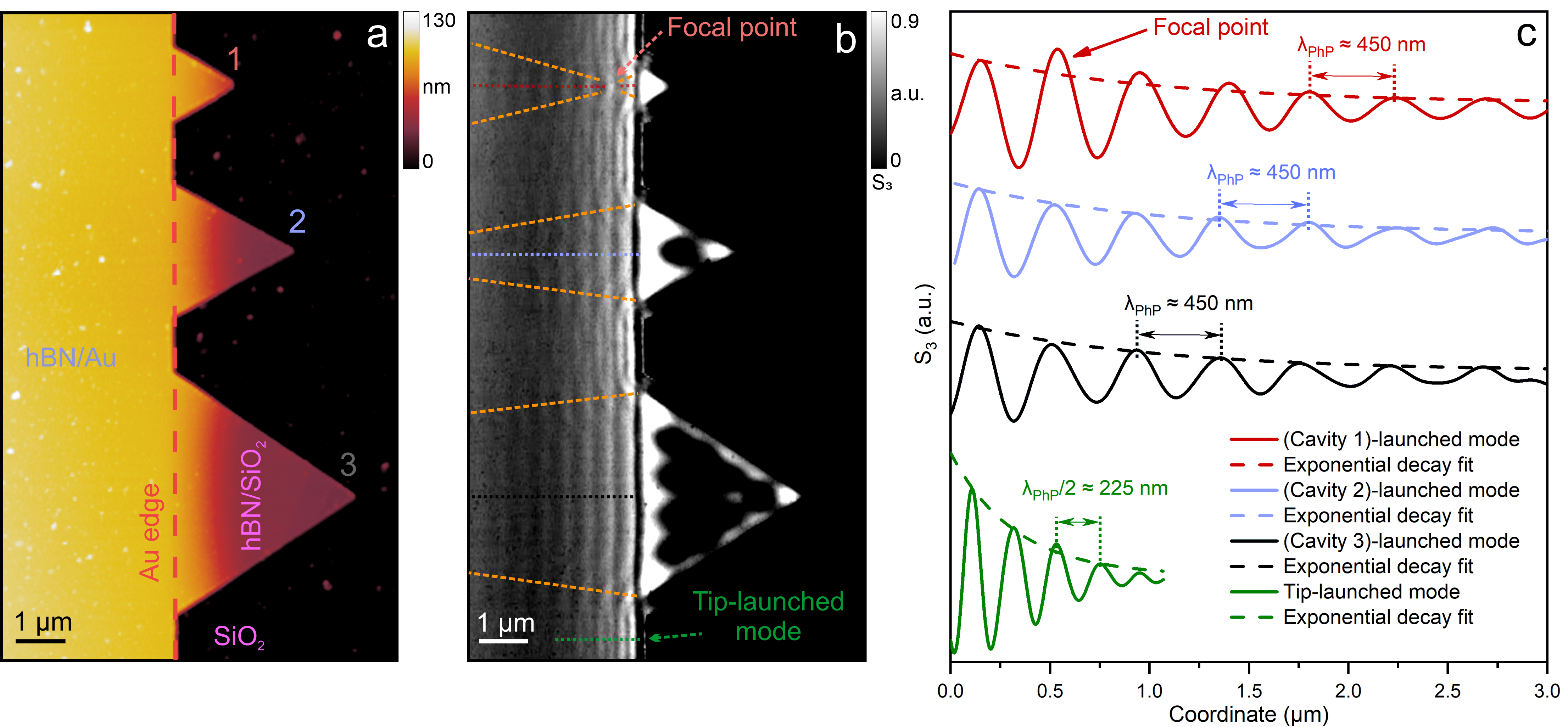}
    \caption{
    	\textbf{Focusing of PhPs using patterned subwavelength cavities.} \textbf{a}) AFM topography of the Au edge with the resonators of triangular shapes. The hBN flake thickness is 45 nm. Red dashed line shows the edge between Au and SiO$_2$. \textbf{b}) Experimental s-SNOM amplitude (S$_3$) map vs. the tip position recorded at $\omega$ = 1450 cm$^{-1}$. \textbf{c}) s-SNOM amplitude (S$_3$) profiles taken across cavity-lanched waves (red, blue, and black lines) and the hBN flake edge on Au (green line). Dashed lines show exponential decay fit for each curve.}
     \label{fig2}
\end{figure*}

Fig.~\ref{fig1}b shows optical amplitude (\textit{s}$_4$) versus tip position on a 42-nm-thick hBN flake placed on top of a 50-nm-thick Au stripe at $\omega$ = 1452 cm$^{-1}$ (see topography and optical phase data in Extended Data Fig.~S1). We focus on three specific regions: i) to the left of the Au stripe, ii) on top of the Au stripe, and iii) to the right of the Au stripe. The first region comprises a triangular hBN piece, where we observe distinct maxima and minima of the optical amplitude (related to the electric field amplitude $(\textrm{Re E}_{\textrm{z}})^2$), characteristic of localized resonant PhP modes\cite{herzig2024high}. In the second region, we identify two types of the PhP waves. The first type appears near the edges of the hBN flake (highlighted by the orange dotted line). The second wave emanates from the cavity (red dotted line). In the third region, additional two PhP waves are visible. The first wave is positioned near the edges of the hBN flake (blue dotted line). The second wave extends from the edge of the Au stripe to the right across the entire hBN bulk (green dotted line). 

The waves, which are localized close to the edge and propagate perpendicularly to it, are the tip-launched PhPs (we do not observe hBN edge-launched fringes in our geometry)\cite{hu2020phonon}. These waves originate from the mid-IR light scattered on the AFM tip. The observed fringes represent only a half-wavelength of the corresponding PhP ($\lambda_{PhP}$/2) since the detection moves along with the excitation \cite{dai2017efficiency, menabde2022image}. Based on this, we attribute the orange and the blue PhP waves to the tip-launched modes on Au and SiO$_2$, respectively. 

Fig.~\ref{fig1}c shows linecuts taken along the dotted lines in the panel Fig.~\ref{fig1}b. As expected, the red and the green curves exhibit doubled wavelengths $\lambda_{PhP}$ (compared to the tip-launched waves on Au and SiO$_2$, respectively). Intriguingly, they also demonstrate notably longer propagation lengths. We attribute these PhPs to the cavity-launched waves \cite{menabde2022image}. In the case of the green curves, the infrared light launches the corresponding PhP in the hBN on SiO$_2$. For the red curves, we assume that the Au edge launches the PhPs inside the cavity where they localize and enhance, and then it leaks back through the semi-transparent interface into the hBN bulk on Au. We emphasize that the tip-launched PhPs (orange and blue curves) only exist when the AFM tip is present in the measurement setup, while the PhPs, indicated by the red and green lines, couple to the far-field light and exist as long as there is far-field light excitation.

Fig.\ref{fig1}d shows a set of zoomed-in maps of the cavity-launched wave at various excitation wavelengths. The launched PhP (white arrow) moves and shrinks along with the resonant maximum position inside the cavity (see the detailed investigation of the dispersion in Extended Data Fig.~S2). Fig.\ref{fig1}e shows optical amplitude profiles taken along the red and the orange lines in Fig.\ref{fig1}d at $\omega$ = 1410 cm$^{-1}$. We find that the electric field is significantly enhanced inside the cavity exhibiting a two-fold larger reflectivity compared to the bare gold signal (and seven-fold larger compared to hBN/SiO$_2$ signal). This observation establishes our subwavelength cavities as effective PhP resonators.

To investigate PhP waves' dependence on their shapes and sizes, we utilize a standard electron beam lithography and plasma etching (see Methods). We fabricate cavities of the desired shapes at the edges of gold patterns. Fig.~\ref{fig2}a shows a topography of one of our samples. Attributed to an exceptional alignment (within $\sim$10 nm) between the e-beam lithography steps, we create the matching edges between the patterned hBN and the Au substrates. This allows us to avoid unwanted leakages of the PhP modes away from our cavities. On the left hand side of the orange dashed line, a 45 nm thick hBN flake covers a 50 nm thick Au stripe. On the right hand side, the patterned hBN triangles hang from the Au edge and lie down on the SiO$_2$ substrate (note a smooth hBN topography height transition across the Au/SiO$_2$ edge). Respectively, Fig.~\ref{fig2}b shows an experimental s-SNOM amplitude (S$_3$) taken at $\omega$ = 1450 cm$^{-1}$ (see its dispersion and the analogous set of data for a 31-nm thick hBN in Extended Data Fig.~S3 and S4, respectively). Inside the cavities, the scattered signal maxima confirm the field localization and enhancement.

\begin{figure*}[t]
    \centering
    \includegraphics[scale=1]{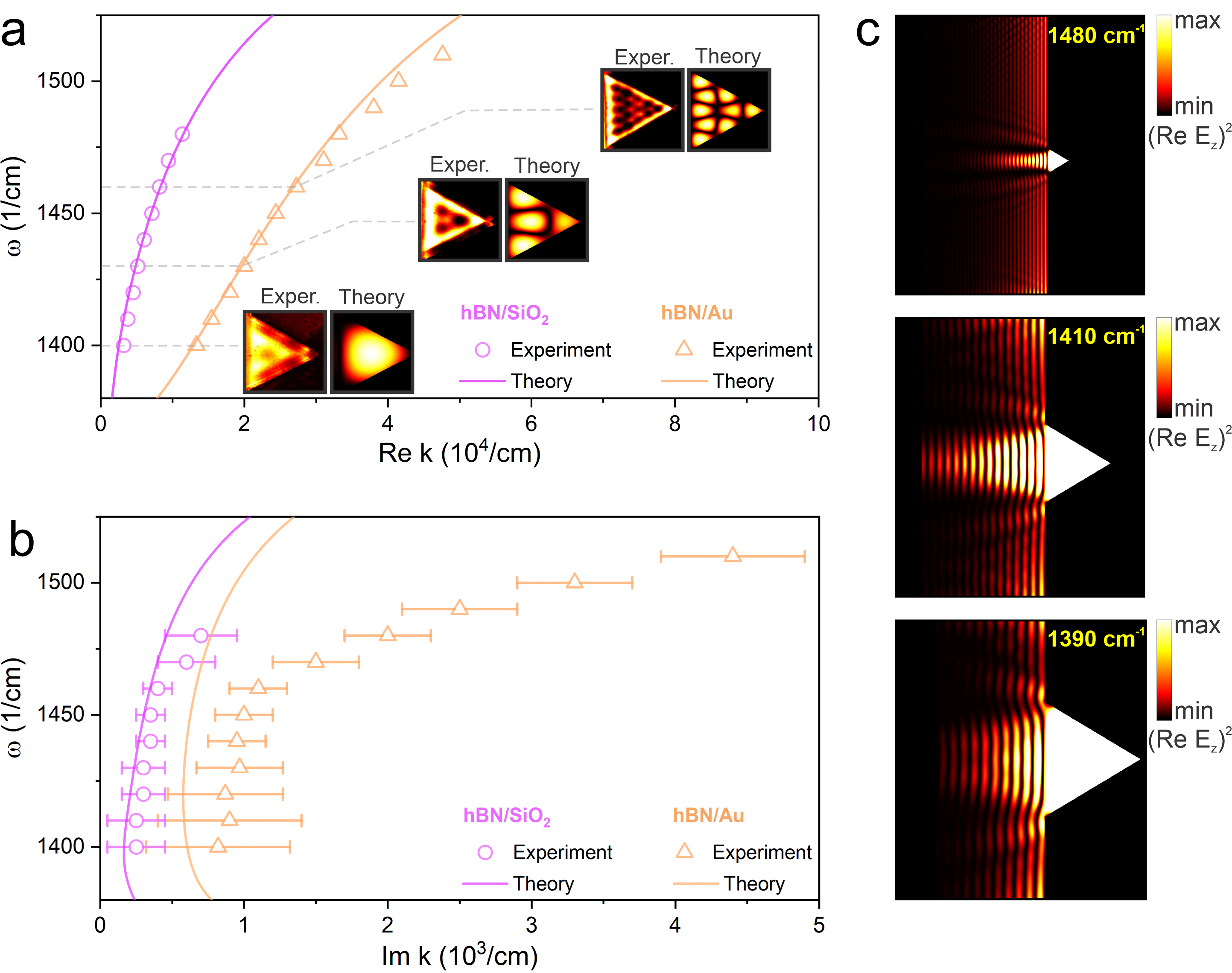}
    \caption{
    	\textbf{Numerical simulations of PhPs alongside comparisons to experimental data.} \textbf{a}) Numerically calculated dispersion of propagating PhP modes generated at the hBN/SiO$_2$ and hBN/Au interfaces (solid lines) compared to the experiment (symbols). Insets show the experimental s-SNOM optical amplitude (S$_4$) maps and the theoretical spatial distributions of $(\textrm{Re E}_{\textrm{z}})^2$ in triangular cavities, for the wavelength indicated by the grey dashed lines. \textbf{b}) Numerically and experimentally obtained dispersion of the attenuation constant ($\textrm{Im k}$) for hBN/SiO$_2$ and hBN/Au interfaces. The error bars indicate the spread of $\textrm{Im k}$ in the experimental data. \textbf{c}) $(\textrm{Re E}_{\textrm{z}})^2$ in triangular cavities which we experimentally probe in Fig.~\ref{fig2}. The distributions are taken at the cavity's resonant frequencies. All calculations are done for hBN thickness of 45~nm.}
     \label{fig3}
\end{figure*}

\begin{figure*}[t]
    \centering
    \includegraphics[scale=0.95]{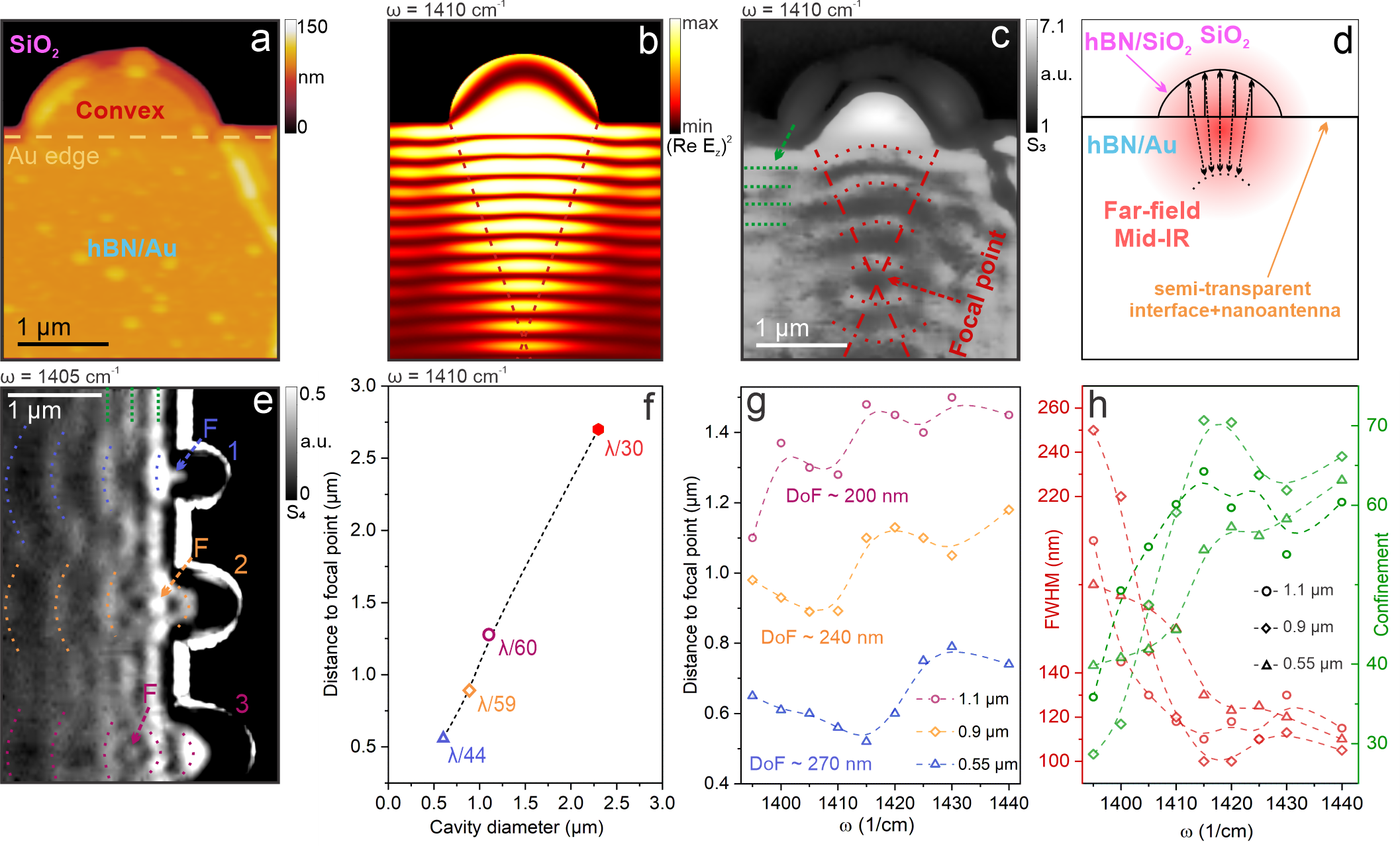}
    \caption{
    	\textbf{Convex-type PhPs focusing.} \textbf{a}) AFM topography of a 2.35-$\mu$m cavity. The light yellow dashed line shows the edge between Au and SiO$_2$. The thickness of hBN is 55 nm. \textbf{b}) Numerical modeling of the cavity. \textbf{c}) Experimental s-SNOM optical amplitude (S$_3$) map recorded at $\omega$ = 1410 cm$^{-1}$. Red dashed lines are a guide for eyes indicating boundaries of cavity-launched polaritons. Dotted lines highlight maxima of optical amplitude for cavity-launched and tip-launched PhPs. Green lines demonstrate tip-launched PhP waves. \textbf{d}) Schematic of the "convex lens"-like cavity operation. \textbf{e}) Experimental s-SNOM optical amplitude (S$_4$) map of three small-diameter cavities (0.55 $\mu$m, 0.9 $\mu$m, and 1.1 $\mu$m, respectively) recorded at $\omega$ = 1405 cm$^{-1}$. \textbf{f}) Dependence of the focal point distance on cavity size at $\omega$ = 1410 cm$^{-1}$, as well as the confinement factor for each point. \textbf{g}) Dependence of the distance to the focal point on the excitation wavelength for small-diameter cavities (Fig.~\ref{fig4}e). The depth of focus (DoF) was determined for each cavity based on the distance oscillations. \textbf{h}) Dependence of a full width at half maximum (FWHM) (red lines) and confinement factor (green lines) on the excitation wavelength for small-diameter cavities.}
     \label{fig4}
\end{figure*}

Similar to the case in Fig.~\ref{fig1}, two different PhP types can be observed. The one that exhibits a $\lambda_{PhP}/2$ distance between intensity maxima is a tip-launched PhP (indicated by a green arrow). This wave results from mid-IR light scattered by the s-SNOM tip, which excites PhPs propagating toward the edge of the sample, reflecting back, and interfering with themselves\cite{dai2017efficiency,dai2014tunable,basov2016polaritons,yoxall2015direct,woo2023selective}. The cavity-launched PhPs show the distance between the intensity maxima of $\lambda_{PhP}$, which contrasts them with the tip-launched PhPs. Remarkably, $\lambda_{PhP}/2$ fringes are completely suppressed in front of the cavities, and only $\lambda_{PhP}$ fringes are observed, which implies a suppressed back-reflection of the tip-launched PhPs at the cavity interface. 

Importantly, we find that the cavity-launched PhPs acquire well-defined focal points. For cavity 1 (see Fig.~\ref{fig2}a,b), the PhP shrinks before and expands after the focal point (marked by the red arrow). The lateral size of the focal point corresponds to a confinement of $\lambda/25$. For cavity 2 and cavity 3, PhPs a focused at focal points outside of the detectable regions. The orange dashed lines serve as guides to the eye, indicating the boundaries of the cavity-launched phonon-polaritons. At the focal point, we identify a "hot spot" that corresponds to the electric field node. This shows that the PhP's width and the focal point location can be effectively tuned by changing the size of a cavity. 

Fig.~\ref{fig2}c shows linecuts taken across the dashed lines in Fig.~\ref{fig2}b (solid lines) and their exponential decay fits (dashed lines). Tip-launched PhPs exhibit rapid amplitude decay because they originate from the interference between tip-launched waves and those reflected from the edge, meaning they must complete a round-trip distance before detection. In contrast, cavity-launched PhPs exhibit oscillations with considerably slower amplitude decay. As a result, the cavity-launched PhPs propagate more than three times farther than those of tip-launched PhPs.

To characterize the decay rate quantitatively, we fit the amplitude of oscillations using an exponentially decaying function $A = A_0\*e^{-\kappa\*x}+A_f$, where $A_0$ and $A_f$ are the fitting parameters related to the initial and the final amplitude values, $x$ is the coordinate of the tip position, and $\kappa$ is the attenuation constant (equals to $2 \times \textrm{Im k}$). The larger the value of $\kappa$, the faster the damping. For the tip-launched PhPs in hBN/Au, we obtain $\textrm{Im k} \sim 3.1\times10^3$ cm$^{-1}$. For the all cavity-launched PhPs, $\textrm{Im k}$ acquires several times smaller values $\textrm{Im k} \sim 1.0\times10^3$ cm$^{-1}$. Thus, the fitting demonstrates that the cavity-launched PhP quenches considerably slower than the ordinary tip-launched PhP. 

\textbf{Numerical simulations}. To shed light on the physics of PhPs in our cavities and to establish an approach to their engineering with the desirable characteristics, we build a theoretical model that predicts optical properties of PhPs and effectively describes the PhP dynamics in the triangular cavities observed in the experiment. Fig.~\ref{fig3}a shows the dispersion of the propagation constant ($\textrm{Re k}$) of the PhP guided modes at the hBN/SiO$_2$ and hBN/Au interfaces. The thickness of the hBN flake is 45~nm. The numerically calculated dispersions appear to be in good agreement with the data extracted from the experiment. The dispersions take into account the waviness of the hBN flake as a 5~nm gap  formed due to the hBN transfer technique. 

Fig.~\ref{fig3}b shows the dispersion of the attenuation constant $\textrm{Im k}$ and compares it with the decay rate  $\kappa=2\times\textrm{Im k}$ obtained from the experiment. The numerical and experimental attenuation constants well-agree with each other between 1400 cm$^{-1}$ and 1460 cm$^{-1}$ but diverge at the frequencies above 1460 cm$^{-1}$. We attribute this divergence to the absorption of residual polymers, specifically PMMA 950 A4 (polymethyl methacrylate, an e-beam lithography resist) and PC (polycarbonate, the top surface of a dry-transfer stamp), remaining on the sample surface after the fabrication process. These materials exhibit high absorption in this spectral range due to the presence of various carbon bonds ~\cite{petibois2009infrared, guo2021effects, tsuda2018spectral, tanaka2015protection}. The dispersion, refractive index and the attenuation constant for the 31- and 42-nm thick hBN can be found in Extended Data Fig.~S5 and S6. We use the complex effective refractive index of PhPs extracted from the dispersions in Fig.~\ref{fig3}a,b to construct a 2D model that captures the PhP wave dynamics in the cavities and waveguides formed at the hBN/SiO$_2$ and hBN/Au interfaces, correspondingly. As can be seen from the insets in Fig.~\ref{fig3}a, the simulated field distributions showcase a reasonable agreement with the experimental optical amplitude maps. The discrepancies between the simulated and experimental distributions, that are observed at the edges of the cavities, are due to the interactions of the s-SNOM probe and edges, while the probe is next to the edge of the cavity~\cite{babicheva2017near}. Notably, this interaction is wavelength dependent. However, in the experiment, the optical amplitude along the Au and hBN edges is noticeably higher, which is attributed to 3D effects at the edges of the heterostructure and buckling of the hBN at the Au edge which cannot be described within our 2D model. Fig.~\ref{fig3}c demonstrates the spatial distributions of the PhPs localized in the triangular hBN/SiO$_2$ cavities and propagating in the hBN/Au waveguide. The distributions show that the triangular cavities can support localized PhP modes that leak into the hBN/Au environment and that the PhP waves are focused at a distance from the triangle base depending on the triangle size, which is also observed in the experiment (see Fig.~\ref{fig2}b).

The schematic analysis of the triangular and rectangular cavities (see Extended Data Fig.~S7) shows that the shape of the PhP's wavefront is closely related to the geometry of the cavity. For example, the size of the triangular cavity determines the focal point, while cavity-launched PhPs of rectangular cavities do not shrink and, as a result, do not obtain a focal point. Conclusively, due to a strong confinement, polaritons obey the laws of the wave optics, which brings up and opportunity to realize "concave lens"-like and "convex lens"-like focusing. 

\textbf{PhP lensing}. Fig.~\ref{fig4}a, \ref{fig4}b, and \ref{fig4}c show the AFM topography, the numerically calculated electric field $(\textrm{Re}, E_{\textrm{z}})^2$, and the experimental near-field optical amplitude (S$_3$) of a 2.35-$\mu$m "convex lens"-like cavity, respectively. This patterned shape exhibits focused PhPs at a point beyond the cavity, similar to triangular cavities. A focal point can be observed where the curvature of the wavefront changes from positive to negative. Fig.~\ref{fig4}d illustrates a schematic of the "convex lens"-like focusing. The PhP waves launched by the Au edge propagate perpendicular to the Au edge inside the cavity and reflect back from the convex cavity edge. Upon reaching the cavity's edge, they reflect normally to the curvature of the reflecting plane. The positive curvature of the reflecting plane results in a positive curvature of the PhP wavefront.

Fig.~\ref{fig4}e shows the s-SNOM optical amplitude (S$_4$) map of three cavities with different diameters: 0.55 $\mu$m, 0.9 $\mu$m, and 1.1 $\mu$m. The focal points are located at varying distances from the cavity/Au interfaces which is determined by its diameter. The focal distance as a function of the cavity diameter for $\omega = 1410$ \textrm{cm}$^{-1}$ is shown in Fig.~\ref{fig4}f. As can be seen, the focal distance linearly depends on the cavity diameter (i.e., on the curvature of the reflective plane). Fig.~\ref{fig4}g demonstrates the distance to the focal point as a function of the frequency of the incident light for three cavities illustrated in Fig.~\ref{fig4}e. The focal distance oscillates with a period specific to each cavity. Based on this, we determine the depth of field (DoF) for these cavities. DoF for classical lenses is the range of distances over which an object remains in focus, given by \( DoF = n\lambda/{(NA)^2} \), where \( n \) is the refractive index, \( \lambda \) is the wavelength, and \( NA \) is the numerical aperture. Each PhP lens has a specific focal length that depends on the geometry of the cavities, but at the same time, DoF for these lenses is finite and depends on the geometry of the cavities as well. By changing the excitation wavelength, we change the real-space PhP field distribution. Therefore, for some wavelengths, the maximum of the PhP field is located exactly in the middle of the focal depth, while for other wavelengths, it is shifted to the edge of field (see the detailed investigation in Extended Data Fig.~S8).

Fig.~\ref{fig4}h demonstrates the full width at half maximum (FWHM) of the focal point as well as the confinement factor calculated as $\lambda / \textrm{FWHM}$ for cavities 1, 2, and 3 at different wavelengths. The FWHM decreases and the confinement increases with the increasing frequency, which is in agreement with the effective refractive index of hBN that grows with frequency (see Extended Data Fig.~S6). A confinement factor as high as 70 can be achieved for the 0.9-$\mu$m cavity at $\omega$ = 1420 \textrm{cm}$^{-1}$, placing it among the highest reported values. 

Using the same approach, we developed a "concave lens"-like cavity (see Extended Data Fig.~S9). The negative curvature of the reflecting plane produces a divergent PhP wavefront and defocusing. Furthermore, multiple cavities launching crisscrossing PhPs can be used for the formation of interference patterns (see Extended Data Fig.~S10), forming a lattice of $(\textrm{Re}, E_{\textrm{z}})^2$ minima and maxima with a period equal to the PhP wavelength at the intersection. Thus, using the interference of cavity-launched PhP with adjusted wavefronts, it is possible to engineer polaritonic lattices for structured light-matter interaction.

\vspace{0.2cm}
\textbf{Discussion.}
To accelerate the technological application of spatially defined PhP modes, it is crucial to develop a method for precisely controlling their spatial distribution and focusing them at specific real-space locations. Significant progress has been made in tuning and focusing PhPs using in-plane-anisotropic hyperbolic materials such as MoO$_3$ \cite{ma2021ghost, chen2020configurable, qu2022tunable}. In the case of uniaxial hyperbolic hBN crystals, most efforts have relied on substrates with in-plane anisotropy in the relevant spectral ranges \cite{chaudhary2019engineering, gong2024dispersion, hajian2020tunable}.

A particularly elegant solution was proposed by Chaudhary et al. \cite{chaudhary2019polariton}, employing a phase-change material (Ge$_3$Sb$_2$Te$_6$ on a CaF$_2$ substrate) to locally modify dielectric constants and focus tip-launched PhPs. However, the influence of such substrates is generally limited, and integrating such uncommon materials into standard silicon-based technology remains challenging. Consequently, the focusing effect remains constrained, and scalable applications are elusive.

One of the most widely adopted methods for focusing PhPs and coupling them to the far-field component of light is the use of metallic scatterers, typically made of gold \cite{huber2008focusing}. This approach is essential because the tip must not be present in operational devices. By employing scatterers with curved surfaces, PhPs can be focused via wave interference. However, this approach has a fundamental limitation: the strength of light coupling to PhP modes depends heavily on the angle between light polarization and the normal to the scatterer surface at the launch point. Consequently, curved scatterers can only efficiently focus PhP modes within a narrow angular range \cite{huber2008focusing}, making them unsuitable for achieving arbitrary PhP lensing in isotropic materials like hBN.

Our approach overcomes these limitations by utilizing standard, technologically accessible materials (Si/SiO$_2$, Ti/Au) as a platform for PhP engineering. We introduce hBN subwavelength cavities at the Au/SiO$_2$ interface, which enable the generation of strongly focused (down to $\lambda/70$) and spatially confined PhPs.  Crucially, our approach resolves the "scatterer" issue mentioned earlier: the Au/SiO$_2$ interface facilitates coupling to far-field light but does not actively participate in the focusing process itself. Instead, the focusing mechanism is governed by the selection of PhPs with specific $k$-vector distributions inside the cavity. As a result, for a fixed Au/SiO$_2$ interface orientation, cavities of varying shapes and sizes can produce different focusing effects (e.g., convex, concave, or straight wavefronts) and different angular ranges of collection. Thus, we believe our approach addresses several persistent challenges in PhP photonics and paves the way for numerous future discoveries.

In conclusion, we have developed hBN sub-wavelength cavities that are compatible with standard lithographic techniques, enabling the launch of focused, long-lived PhPs. We fabricate the subwavelength hBN cavities at the interfaces between Au and SiO$_2$ partially transparent for PhPs. The localized modes exhibit a significant enhancement of the electric field inside the cavities, launching PhPs that remain spatially confined, with their confinement determined by the shape and size of the cavities. Using s-SNOM, we studied the real-space optical contrast of the cavity-launched PhPs. Based on the PhP's wavelength, we demonstrated coupling to the far-field component of the mid-IR light. The far-field component, which is resonantly enhanced in the cavities, acquires sufficient momentum to launch polaritons into the bulk of the hBN/Au heterostructures through the scattering process at the SiO$_2$/Au interfaces. The shape of our resonators determines both the focal point location and the wavefront geometry. By creating different forms of cavities, we showed a focusing effect similar to a "concave lens" and a "convex lens." In particular, the "convex lens"-like cavities enable field confinement down to $\lambda/70$ at the focal point. By this, we realize a novel technique for generating and focusing spatially confined in-plane PhPs compatible with the standard fabrication approaches.


\section*{\textsf{Methods}}
\small
\subsection*{Device fabrication.}
\noindent
Si$^{++}$/SiO$_2$ substrates were pre-patterned using standard photolithography and metal deposition procedures. For photolithography, SUSS MicroTec MJB4 mask aligner and SPR-700 photoresist were used. After the photoresist development, we deposited metal (Ti/Au, 10/50 nm) using the Oerlikon 450B electron beam vacuum deposition system.

hBN flakes were prepared via mechanical exfoliation using PDMS adhesive tapes and were transferred on Si/SiO$_2$ substrates for selection. Relying on optical contrast, we selected flakes of suitable thicknesses and transferred them on the pre-patterned Si$^{++}$/SiO$_2$ substrates using the PMDS/PC dry transfer method\cite{wang2013one,purdie2018cleaning}.

For the cavity fabrication, we used the Raith EBPG5200 electron beam lithographer and PMMA 950 A4 e-beam resist. After the resist development, exposed hBN was etched out using Oxford ICP-RIE plasma etcher with CHF$_3$ plasma source (40 sscm, 90 mTorr, 60 W).

\subsection*{Near-field optical measurements.}
\noindent

We used a commercially available scattering-type scanning near-field microscope (s-SNOM) developed by Neaspec/Attocube (cryo-neaSCOPE) with a pseudo-heterodyne mode to a detect a background-free optical response ($\Omega \pm NM$) using the decoupled optical amplitude and phase\cite{ocelic2006pseudoheterodyne, moreno2017phase,keilmann2004near}. PtIr-coated AFM tips (Nanoworld, 23 nm coating) with a resonant frequency $\approx$ 250 kHz were used to obtain the topography and the optical maps. AFM tips were illuminated by a tunable commercially available mid-infrared quantum cascade laser (QCL) using 1-2 mW of power (MIRcat, $\lambda$ = 5.5-11.5 $\mu$m from the DRS Daylight Solutions Inc.) All optical measurements presented in the manuscript were obtained on the 3rd or the 4th harmonics (i.e., S$_3$ and S$_4$, respectively) to ensure the background-free detection.

\subsection*{Numerical model}
\noindent

To describe the optical phenomena in PhPs at the hBN/SiO$_2$ and hBN/Au interfaces, we employ numerical calculations in COMSOL Multiphysics, Wave Optics module. The dispersions of the propagation and attenuation constants of the PhP modes alongside the complex effective refractive indices of these modes are calculated in a 4-layer structure, where the top layer is presented by air, the second layer is presented by hBN which is an anisotropic (along the hard axis) material with permittivity tensor taken from~\cite{kumar2015tunable} and optical axis directed perpendicular to the layers, the third layer is a narrow air gap, and the fourth layer is either SiO$_2$ or Au with refractive indices taken from \cite{kischkat2012mid,babar2015optical}. The obtained effective refractive indices are used as refractive indices for hBN/SiO$_2$ and hBN/Au domains in the 2D model, and the surrounding space is treated as a dielectric with a refractive index of 1. As a source for the 2D model, we specify the surface current density at the edges of the hBN/SiO$_2$, hBN/Au domains, and the surrounding space. 

\bibliographystyle{naturemag-sergi}
\bibliography{hBN.bib}

\vspace*{-0.2cm}
\section*{\textsf{Acknowledgements}}
\vspace*{-0.2cm}
\noindent
We thank Andrey Bogdanov, Lorenzo Orsini, Frank H. L. Koppens and Hanan Herzig Sheinfux for helpful discussions. The authors thank Morten R. Eskildsen for his assistance in setting up the experiment. P.S. acknowledges support from the start-up fund provided by the University of Notre Dame (\#373837). K.W. and T.T. acknowledge support from the JSPS KAKENHI (Grant Numbers 21H05233 and 23H02052) and World Premier International Research Center Initiative (WPI), MEXT, Japan.

\vspace*{-0.2cm}
\section*{\textsf{Author contributions}}
\vspace*{-0.2cm}
{\small
\noindent
P.S. and B.B conceived and designed the experiment. B.B. fabricated samples and performed the measurements. The data were analyzed and interpreted by B.B., S.L. and P.S. The hBN crystals were provided by K.W. and T.T. S.L performed the numerical calculations. B.B, S.L., and P.S. wrote the manuscript. P.S. supervised the work.
}

\vspace*{-0.2cm}
\section*{\textsf{Competing Financial Interests}}
\vspace*{-0.2cm}
{\small
\noindent
The authors declare no competing financial interests.
}

\vspace*{-0.2cm}
\section*{\textsf{Data Availability Statement}}
\vspace*{-0.2cm}
{\small
\noindent
The data that support the findings of this study is available from the corresponding authors upon reasonable request.
}
\newpage
\end{document}


\title{\Large\textsf{\papertitle}}

\author{Bogdan Borodin}
\email{bborodin@nd.edu}
\affiliation{\footnotesize Department of Physics and Astronomy, University of Notre Dame, Notre Dame, IN 46556,~USA}
\affiliation{\footnotesize Stavropoulos Center for Complex Quantum Matter, University of Notre Dame, Notre Dame, IN 46556,~USA}

\author{Sergey Lepeshov}
\affiliation{Department of Electrical and Photonics Engineering, DTU Electro, Technical University of Denmark, Building 343, DK-2800 Kgs.~Lyngby, Denmark.}
\affiliation{NanoPhoton - Center for Nanophotonics, Technical University of Denmark, Ørsteds Plads 345A, DK-2800 Kgs.~Lyngby, Denmark.}

\author{Kenji Watanabe}
\affiliation{\footnotesize Research Center for Functional Materials, National Institute for Materials Science, 1-1 Namiki, Tsukuba 305-0044,~Japan}

\author{Takashi Taniguchi}
\affiliation{\footnotesize International Center for Materials Nanoarchitectonics, National Institute for Materials Science,  1-1 Namiki, Tsukuba 305-0044,~Japan}

\author{Petr Stepanov}
\email{pstepano@nd.edu}
\affiliation{\footnotesize Department of Physics and Astronomy, University of Notre Dame, Notre Dame, IN 46556,~USA}
\affiliation{\footnotesize Stavropoulos Center for Complex Quantum Matter, University of Notre Dame, Notre Dame, IN 46556,~USA}

\keywords{phonon-polariton, polaritonic optics, strong confinement, hexagonal boron nitride, focusing}


\begin{abstract}
\vspace*{0.2cm}

\end{abstract}

\maketitle

\renewcommand{\figurename}{Fig.}
\setcounter{equation}{0}
\setcounter{figure}{0}
\setcounter{table}{0}
\makeatletter
\renewcommand{\theequation}{S\arabic{equation}}
\renewcommand{\thefigure}{S\arabic{figure}}
\renewcommand{\thetable}{\arabic{table}}
\renewcommand{\bibnumfmt}[1]{[#1]}
\renewcommand{\citenumfont}[1]{#1}

\onecolumngrid

\begin{figure}[H]
    \centering
    \includegraphics[scale=1.0]{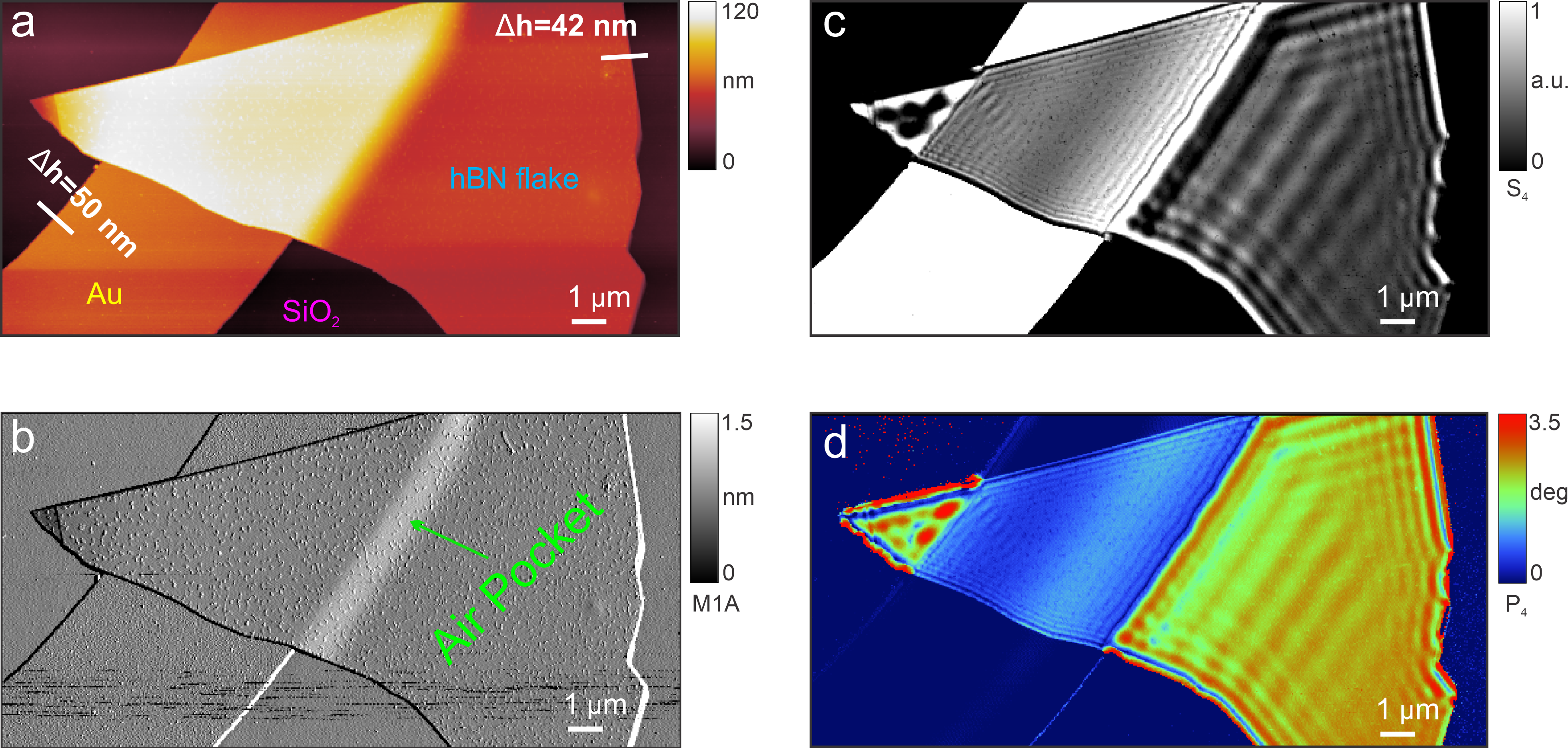}
    \caption{\textbf{Additional data on the as-transferred hBN flake on Au.} \textbf{a}) AFM topography map. \textbf{b} Mechanical amplitude map. \textbf{c}) Optical amplitude (S$_4$) map at $\omega$ = 1450 cm$^{-1}$. \textbf{d}) Optical phase (P$_4$) map at $\omega$ = 1450 cm$^{-1}$.}
    \label{fig:intro}  
\end{figure}

\newpage
\begin{figure}[H]
    \centering
    \includegraphics[scale=0.9]{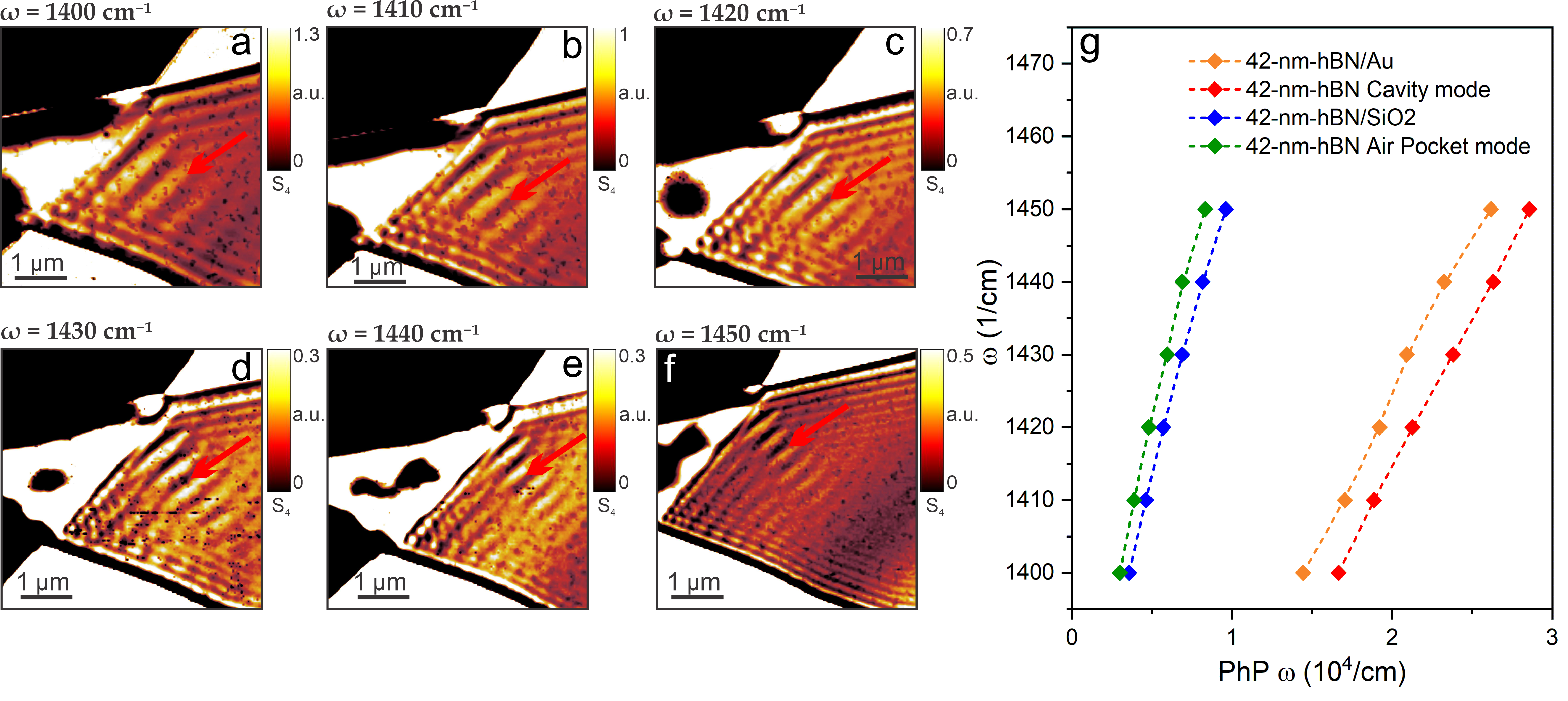}
    \caption{\textbf{PhP dispersion in naturally fabricated hBN cavity.} \textbf{a-f}) Optical amplitude (S$_4$) maps at various excitation $\omega$ = 1400-1450 cm$^{-1}$. The red arrow shows the cavity-launched mode. \textbf{g}) Dispersion relations of different phonon-polariton modes. The green and blue lines correspond to the modes shown in Fig. 1b of the main text. }
    \label{fig:intro}  
\end{figure}

\newpage
\begin{figure}[H]
    \centering
    \includegraphics[scale=0.95]{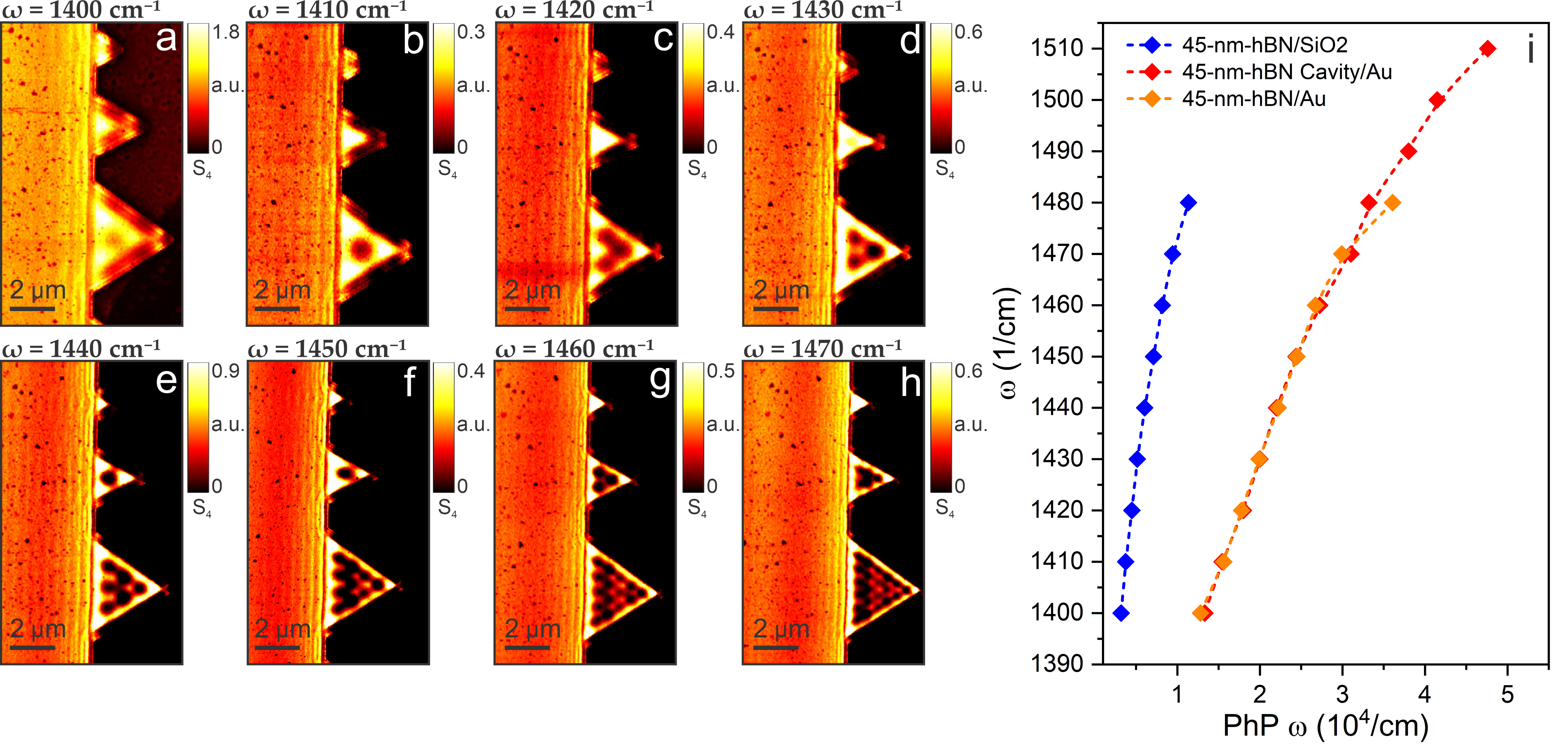}
    \caption{\textbf{PhP dispersion in 45-nm-thick hBN cavities.} \textbf{a-h}) Optical amplitude (S$_4$) maps at various excitation $\omega$ = 1400-1470 cm$^{-1}$ for 45-nm-thick cavities. \textbf{i}) Dispersion of phonon-polariton modes for the 45-nm-thick flake and cavities.}
    \label{fig:intro}  
\end{figure}

\newpage
\begin{figure}[H]
    \centering
    \includegraphics[scale=0.95]{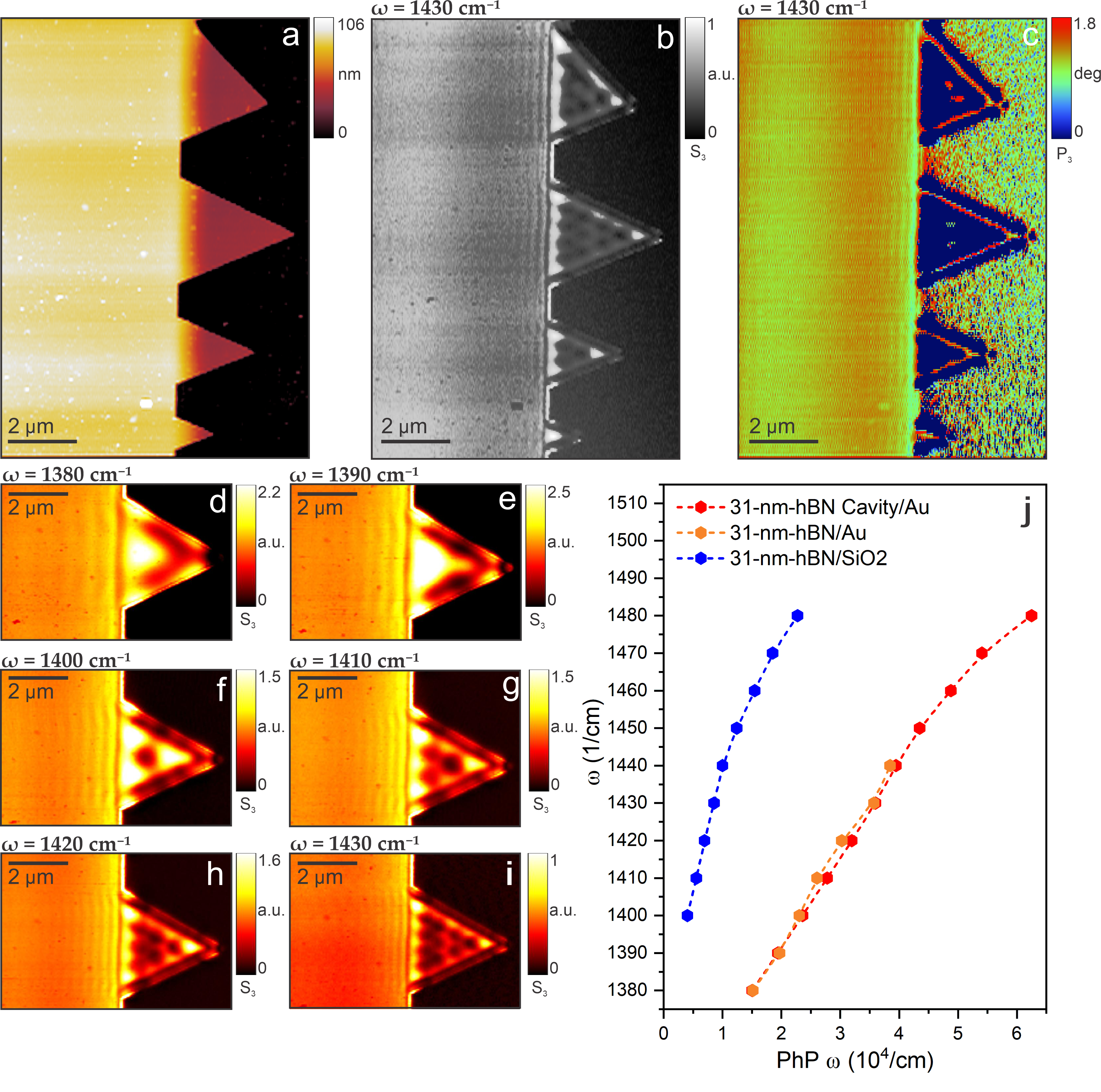}
    \caption{\textbf{31-nm-thick hBN flake and cavities.} \textbf{a}) AFM topography of the cavities. \textbf{b}) Optical amplitude (S$_3$) map at $\omega$ = 1430 cm$^{-1}$. \textbf{c}) Optical phase (P$_3$) map at $\omega$ = 1430 cm$^{-1}$. \textbf{d-i}) Optical amplitude (S$_3$) maps at various excitation $\omega$ = 1380-1430 cm$^{-1}$ for the second from the top cavity. \textbf{j}) Dispersion of the phonon-polariton modes in the 31-nm-thick flake and cavities.}
    \label{fig:intro}  
\end{figure}

\newpage
\begin{figure}[H]
    \centering
    \includegraphics[scale=0.2]{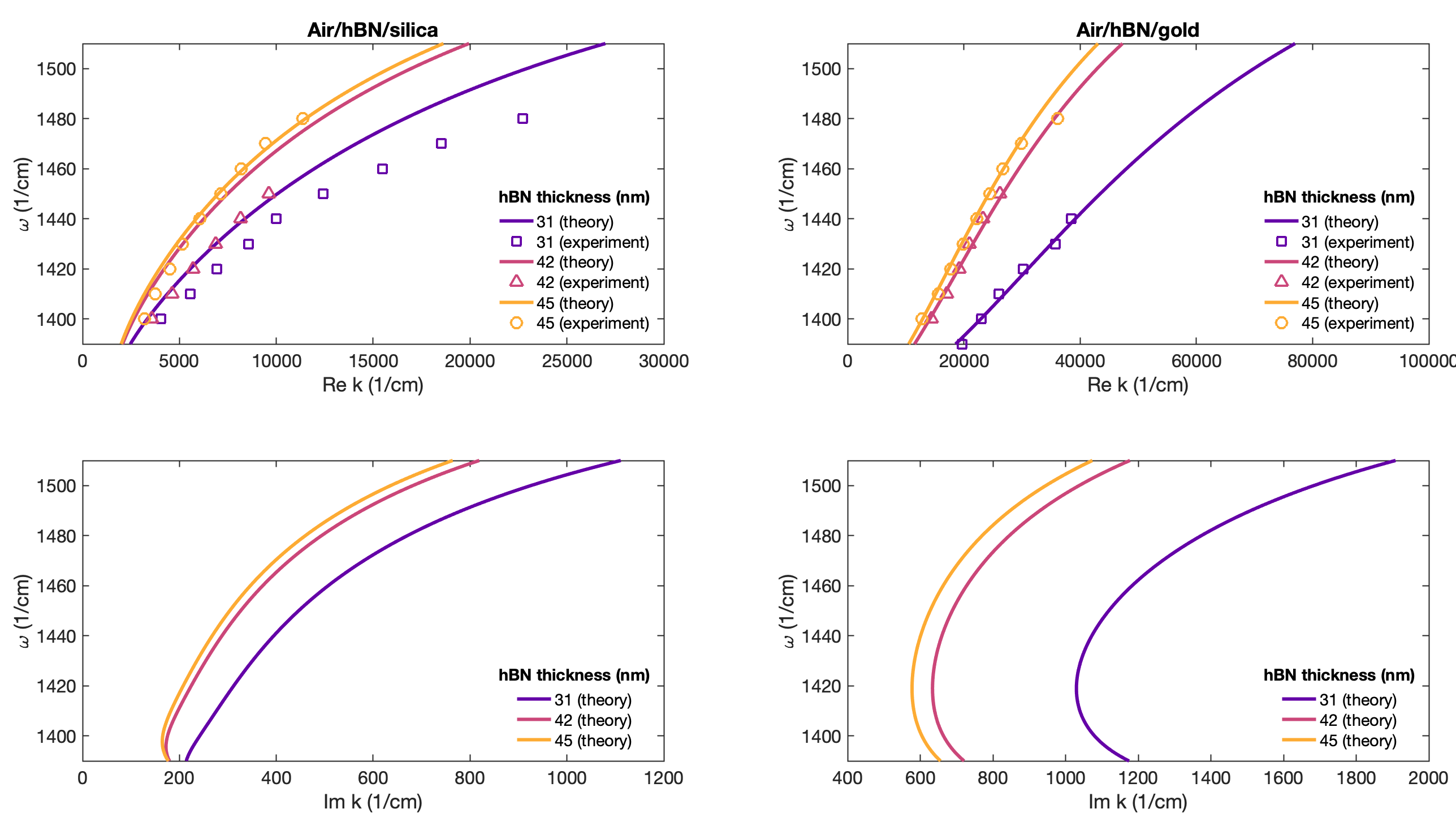}
    \caption{\textbf{Additional data on comparison between the experiment and numerical simulations.} \textbf{a, b})Comparisons between theoretical and experimental dispersions of propagation constants of the PhPs at hBN/SiO$_2$ and hBN/Au interfaces for the hBN thicknesses of 31, 42, and 45~nm. \textbf{c, d}) Dispersions of attenuation constants for the same interfaces. }
    \label{fig:intro}  
\end{figure}

\newpage
\begin{figure}[H]
    \centering
    \includegraphics[scale=0.2]{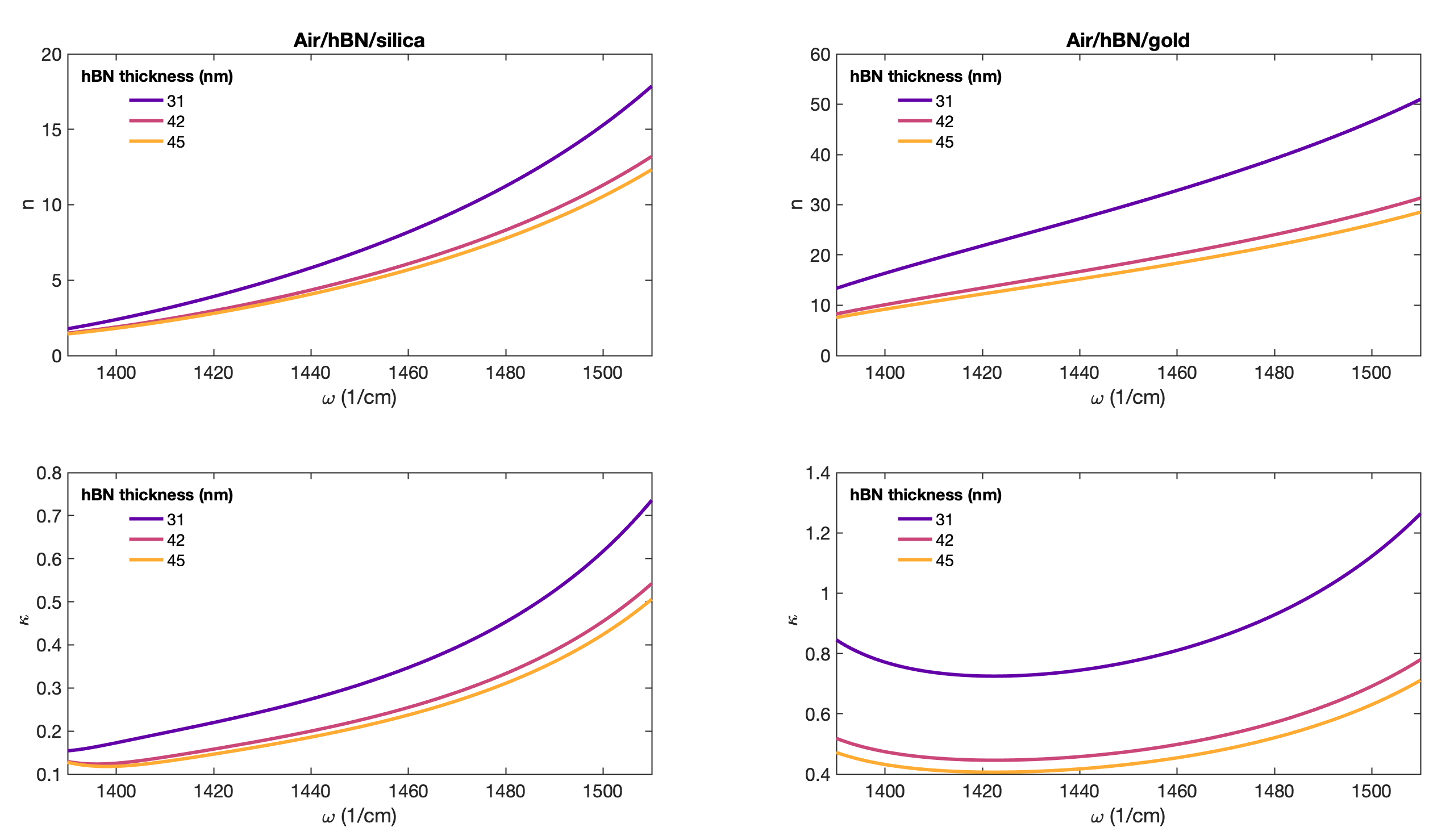}
    \caption{\textbf{Theory.} \textbf{a, b}) Refractive indexes ($\textrm{n}$)  of the PhPs at hBN/SiO$_2$ and hBN/Au interfaces for the hBN thicknesses of 31, 42, and 45~nm. \textbf{c, d}) Attenuation constants ($\kappa$) for the same interfaces. }
    \label{fig:intro}  
\end{figure}

\newpage
\begin{figure}[H]
    \centering
    \includegraphics[scale=1]{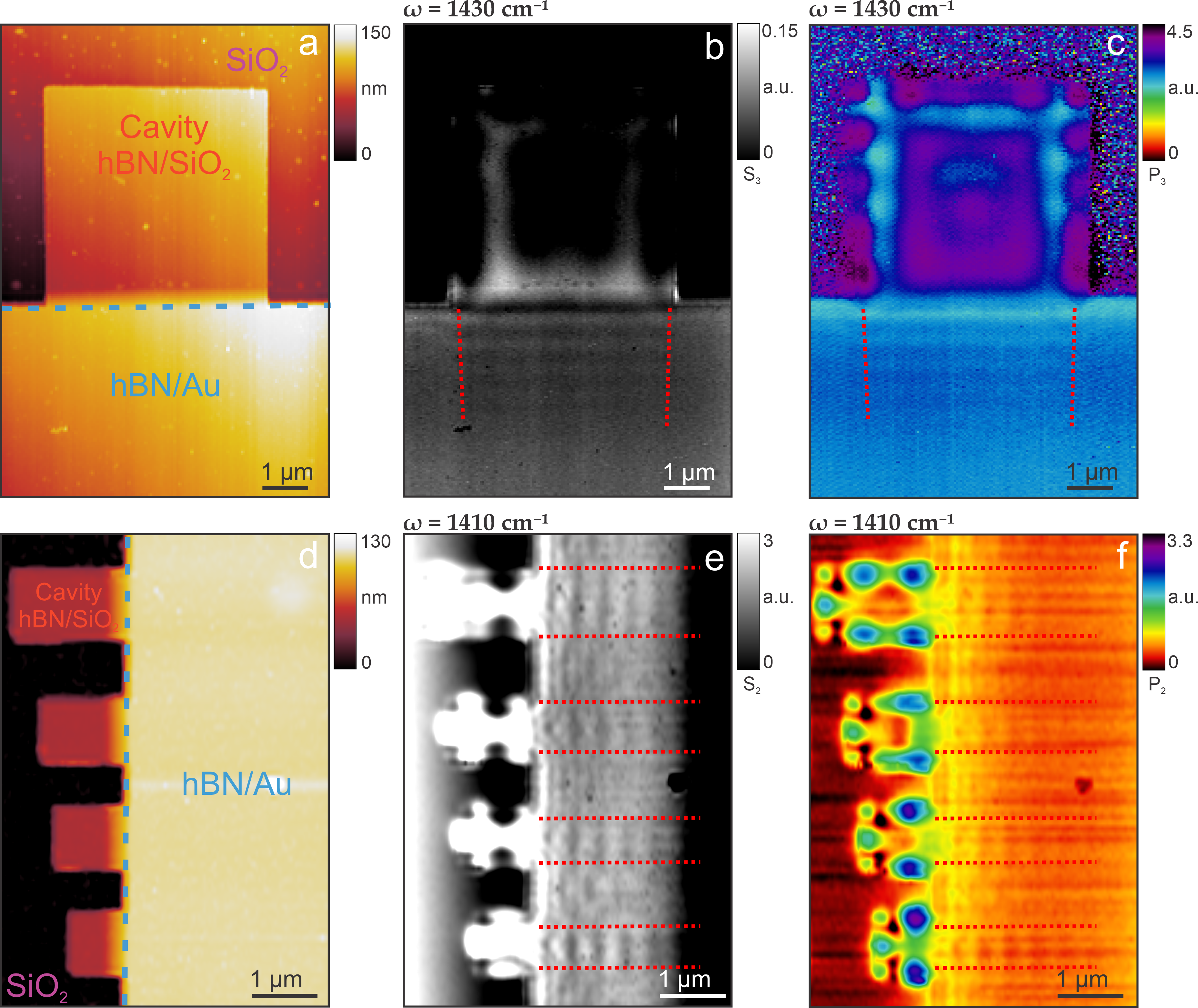}
    \caption{\textbf{Rectangular cavities.} \textbf{a}) AFM topography of a large 31-nm-thick cavity. \textbf{b}) The corresponding optical amplitude (S$_3$) map at $\omega$ = 1430 cm$^{-1}$. \textbf{c}) Optical phase (P$_3$) map at $\omega$ = 1430 cm$^{-1}$. \textbf{d}) AFM topography of small 55-nm-thick cavities. \textbf{e}) Optical amplitude (S$_2$) map at $\omega$ = 1410 cm$^{-1}$. \textbf{f})  Optical phase (P$_2$) map at $\omega$ = 1410 cm$^{-1}$.}
    \label{fig:intro}  
\end{figure}

\newpage
\begin{figure}[H]
    \centering
    \includegraphics[scale=1]{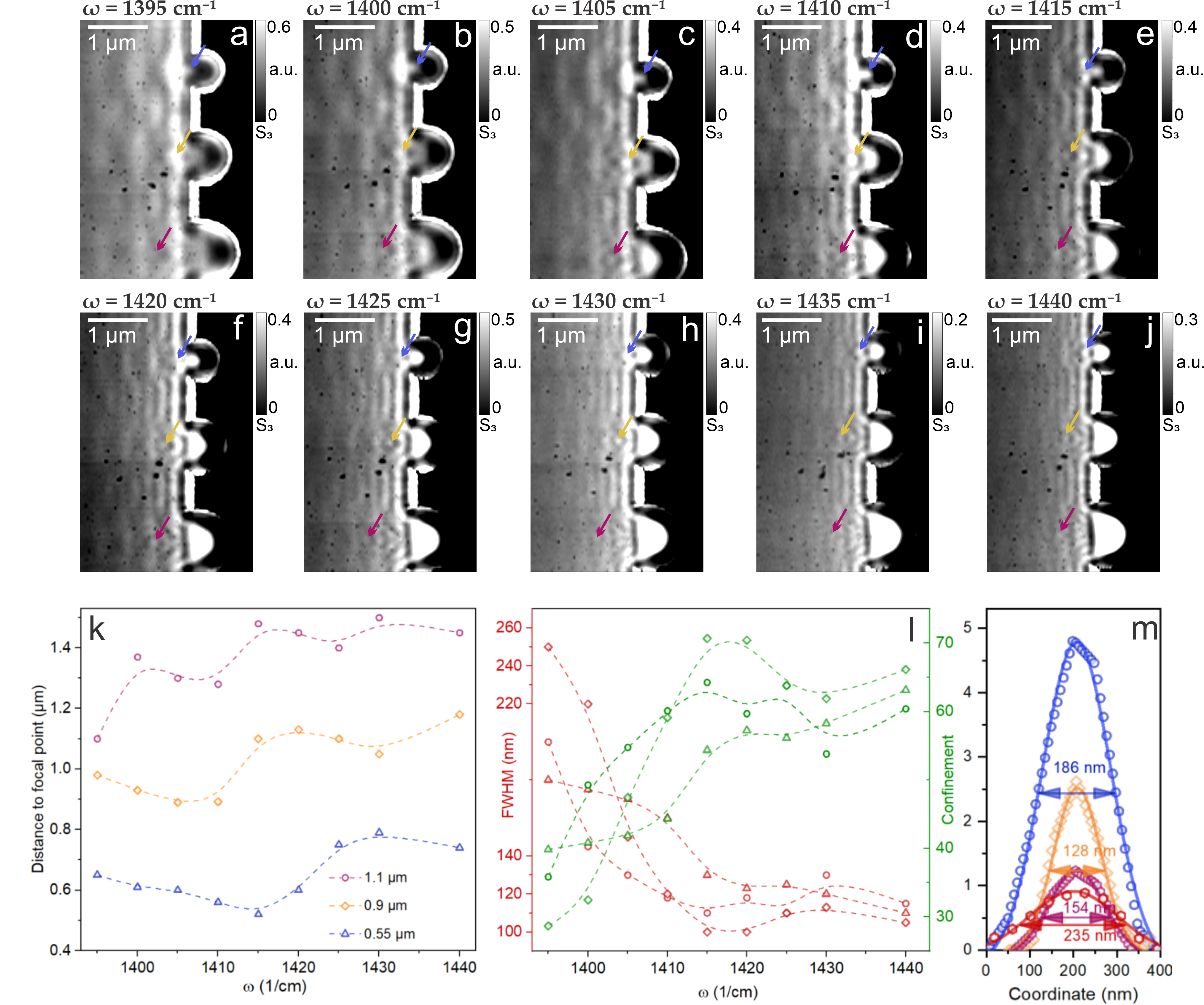}
    \caption{\textbf{a–j)} Experimental s-SNOM optical amplitude (S$_4$) maps of three small-diameter cavities (0.55~$\mu$m, 0.9~$\mu$m, and 1.1~$\mu$m, respectively) recorded at $\omega = 1395$–$1440$~cm$^{-1}$.  \textbf{k)} Dependence of the distance to the focal point on the excitation wavelength for small-diameter cavities.  \textbf{l)} Dependence of the full width at half maximum (FWHM) (red lines) and the confinement factor (green lines) on the excitation wavelength for small-diameter cavities.  \textbf{m)} An example of focal point characterization.}
    \label{fig:intro}  
\end{figure}

\newpage
\begin{figure}[H]
    \centering
    \includegraphics[scale=1]{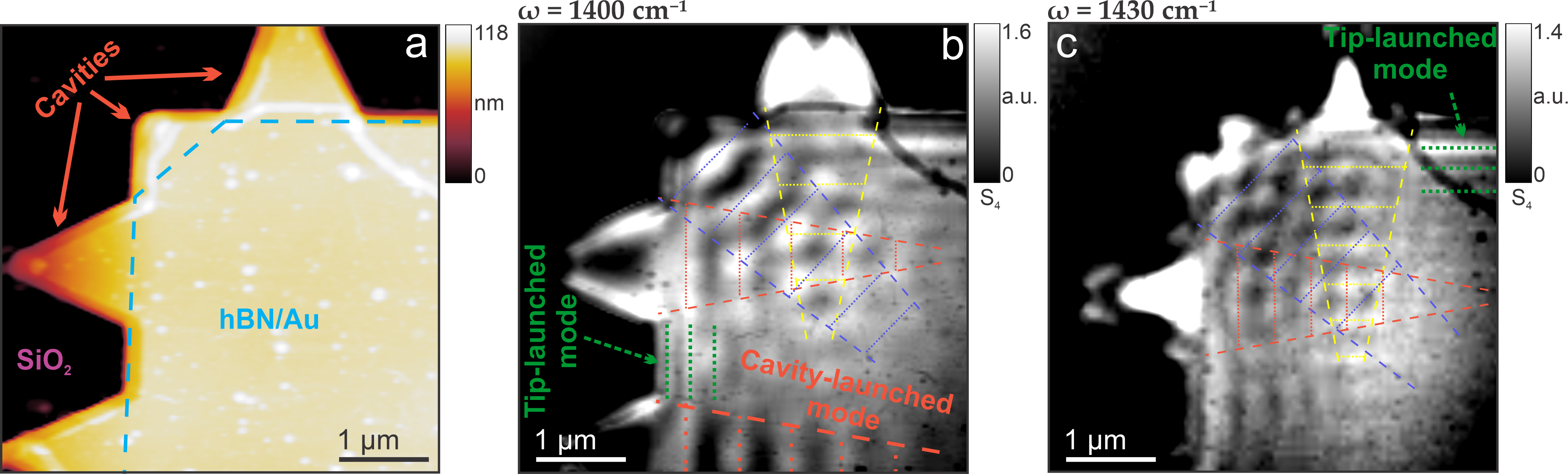}
    \caption{\textbf{55-nm-thick triangular cavities for PhPs interference.} \textbf{a}) AFM topography of the cavities. \textbf{b}) Optical amplitude (S$_4$) map at $\omega$ = 1400 cm$^{-1}$. \textbf{b}) Optical amplitude (S$_4$) map at $\omega$ = 1430 cm$^{-1}$.}
    \label{fig:intro}  
\end{figure}
